\documentstyle[12pt]{article}

\newcommand{\E}{{\cal E}}
\newcommand{\e}{{\rm\bf e}}
\newcommand{\be}{{\rm\bf e}_{\;\mbox{\scriptsize bare}}}
\renewcommand{\j}{j^{\alpha}}
\newcommand{\bj}{j^{\alpha}{}_{\mbox{\scriptsize bare}}}
\newcommand{\r}{r_{\mbox{\scriptsize min}}}
\renewcommand{\t}{t_{\mbox{\scriptsize start}}}
\begin{document}
\hspace{9cm}\vbox{\hbox{IASSNS-HEP-99/42}
\hbox{hep-th/9906241}}
\vspace{1cm}
\begin{center}
{\LARGE\bf The vacuum backreaction on a\\ pair creating source}
\end{center}
\begin{center}
{\bf G.A. Vilkovisky}
\end{center}
\begin{center}
Lebedev Physics Institute and Research Center in Physics,\\
Leninsky Prospect 53, Moscow 117924, Russia
\end{center}
\vspace{2cm}
\begin{abstract}
Solution is presented to the simplest problem about the vacuum
backreaction on a pair creating source. The backreaction effect
is nonanalytic in the coupling constant and restores completely
the energy conservation law. The vacuum changes the {\it kinematics}
of motion like relativity theory does and imposes a new upper bound
on the velocity of the source.
\end{abstract}

\newpage

The phenomenon of the vacuum instability caused by nonstationary
external fields received much attention owing to its significance
in the black hole physics (see the references in [1]) but the
phenomenon itself is quite general. A charged source of a nonstationary
field is capable of creating from the vacuum real particles having
the same type of charge (electrical, gravitational, etc.). When the
frequency of the source exceeds the threshold of pair creation, it 
emits a flux of energy and charge carried by the created particles.
However, the problem with external field is physically incomplete
since it does not answer the question where the energy of the
created particles comes from. It is clear that the vacuum particle
production is only a mechanism of the energy transfer. The energy
comes ultimately from the source of the external field, and there
emerges a question: how much energy can be extracted from a source
through the vacuum mechanism?

An attempt to answer this question without taking into account
the backreaction of the vacuum on the motion of the source leads
only to a contradiction with the energy conservation law. The
radiation of black holes is only one (although the most glaring)
example. Typically, the radiation rate grows unboundedly with
the energy of the source, and, at a sufficiently high energy,
the source appears to give more than it has. This is the case
even in QED [1,2].

Of course, one expects that the corrections stemming from the
self-consistent equations for the expectation values of the field
will remove the contradiction but one should realize that the
vacuum radiation is a purely quantum effect, and, therefore,
a quantum correction to the external field will result only in
a higher-order correction to the radiation energy. The backreaction
effect capable of restoring the energy conservation law can only
be nonanalytic in the coupling constant.

Below I present the solution of the self-consistent problem for the
simplest model of a pair creating source. The question receives an
answer but the significance of this answer seems to surpass the
significance of the question.

The model is an electrically charged spherical shell expanding
in the self field. Below, $r=\rho (t)$ is the law of expansion of 
the shell, $e$ and $M$ are the full charge and mass of the shell, 
and $\E$ is its energy in excess of the rest energy ($c=1$).
It is assumed that before some time instant $t=\t$ the shell was
kept at the state of maximum contraction $r=\r$ and next was let go.
The world line of the shell is shown in Fig. 1.

Since the shell moves with acceleration, it creates particles from
the vacuum. It radiates at a short stage of its evolution near
$t=\t$ where the acceleration is maximum. The bigger the energy $\E$,
the bigger is this acceleration, and the more violent is the creation
of particles. Therefore, it is interesting to consider the
ultrarelativistic shell $(\E/M)\gg 1$.

Without predetermining the law of motion $\rho (t)$, one may assume
that the shell expands monotonically with an increasing velocity
which at $t=\infty$ reaches some finite value ${\dot\rho}(\infty)$.
Then ${\dot\rho}(\infty)$ may serve as a measure {\it at late time}
of the acceleration at $\t$. As $(\E/M)\to\infty$, the velocity
${\dot\rho}(t)$ approaches $1$ at all $t$ except in a small sector
near $t=\t$. The world line of the shell approaches then the
broken line in Fig. 1. These assumptions are valid for the classical
motion of the shell, and they cannot be invalidated by the quantum
corrections since these corrections are small.

Let $\Delta\E$ and $\Delta e$ be the energy and charge emitted by
the shell for the whole its history. Using the methods of Refs. [1,2],
these quantities can be calculated with an arbitrary law of motion
$\rho (t)$. For the ultrarelativistic motion under the assumptions above
the result is\footnote{In the high-frequency approximation [2] which
in the present case is provided by the condition $(\E/M)\gg 1$,
the radiation flux does not depend on the mass of the vacuum particles.}
\begin{equation}
\frac{\Delta\E}{\E}=\frac{\Delta e}{e}=-\frac{\kappa^2}{24\pi}
\log\sqrt{1-{\dot\rho}(\infty)}+\kappa^2 O(1)\; ,
\qquad {\dot\rho}(\infty)\to 1
\end{equation}
where $O(1)$ denotes the terms that remain finite as 
${\dot\rho}(\infty)\to 1$, and $\kappa^2 >0$ is the constant of coupling
of the electromagnetic field of the shell to the vacuum charges
(8 times the fine structure constant for the electron-positron vacuum).
Inserting in (1) the ${\dot\rho}(\infty)$ calculated from the
{\it classical} law of motion
\begin{equation}
\frac{M}{\sqrt{1-{\dot\rho}^2}}+\frac{1}{2}\frac{e^2}{\rho}=
M+\E
\end{equation}
one obtains the result
\begin{equation}
\frac{\Delta\E}{\E}=\frac{\Delta e}{e}=\frac{\kappa^2}{24\pi}
\log\frac{\E}{M}\; ,\qquad \frac{\E}{M}\to\infty
\end{equation}
which manifestly contradicts the energy conservation law.

The self-consistent problem to be solved for obtaining the correct
result is as follows. Any spherically symmetric electromagnetic
field is determined by a single function $\e (t,r)$ which is the
charge contained at the time instant $t$ inside the sphere of
radius $r$. In terms of this function, the electric field $E$
and electromagnetic current $\j$ are respectively of the form
\begin{equation}
E=\frac{\e (t,r)}{r^2}\qquad ,\qquad -4\pi r^2\j=
\Bigl(\nabla^{\alpha}r\frac{\partial}{\partial t}+
\nabla^{\alpha}t\frac{\partial}{\partial r}\Bigr)\e (t,r)\; .
\end{equation}
The $\e (t,r)$ must satisfy the condition of regularity of the
electric field at $r=0\,$, $\e (t,0)=0$, and the normalization
condition $\e (t,\infty)=e$. With these boundary conditions one is
to solve the expectation-value equations
\begin{equation}
\j+\frac{\kappa^2}{2\pi}\gamma (-\Box)\j=\bj \quad ,
\end{equation}
\begin{equation}
\gamma (-\Box)=\frac{1}{12}\biggl(\int\limits_{4m^2}^{\infty}
d\mu^2\, \Bigl(1-\frac{4m^2}{\mu^2}\Bigr)^{3/2}
\frac{1}{\mu^2-\Box}-\int\limits_0^{\infty}d\mu^2\,
\frac{1}{\mu^2+m^2}+\frac{8}{3}\biggr)
\end{equation}
with the retarded resolvent $1/(\mu^2-\Box)$ [1,2]. In (5)-(6),
$m$ is the mass of the vacuum particles, and $\bj$ is expressed
through
\begin{equation}
\be (t,r)=e\,\theta\left(r-\rho (t)\right)
\end{equation}
by the same formula as in (4). The set of equations is closed by
adding the equation of motion of the shell
\begin{equation}
\frac{d}{dt}\biggl(\frac{M{\dot\rho}}{\sqrt{1-{\dot\rho}^2}}\biggr)
=e\,\frac{{\e}_+ (t)+{\e}_- (t)}{2\rho^2}
\end{equation}
where, as appropriate for the charged surface, the force exerted
on the shell is determined by one half of the sum of the electric
fields on both sides of the shell:
\begin{equation}
{\e}_{\pm}(t)=\e (t,\rho (t)\pm 0)\quad .
\end{equation}

Reserving the procedure of solving for an extended publication,
I shall present only the final result. The solution for the
force in (8) is
\begin{equation}
{\e}_+ (t)+{\e}_- (t)=e+e\,\frac{\kappa^2}{24\pi}\log\left(1-{\dot\rho}^2
(t)\right)+\kappa^2 O(1)\quad .
\end{equation}
Since $\kappa^2$ is small, and $O(1)$ is uniformly bounded, the term
$\kappa^2 O(1)$ in (10) can be discarded. The force of the vacuum 
backreaction depends on the velocity. Nevertheless, the equation of
motion (8) with this force admits the energy integral:
\begin{equation}
M\int\limits_{\textstyle 1}^{\textstyle 1/\sqrt{1-{\dot\rho}^2}}
\frac{dx}{\displaystyle 1-\frac{\kappa^2}{12\pi}\log x}+
\frac{1}{2}\frac{e^2}{\rho}=\E
\end{equation}
which at $\kappa^2 =0$ goes over into the classical law (2). There is
no problem with the singularity of the integral in (11). It is never
reached. Like in (2), for a given energy, the velocity ${\dot\rho}$
reaches its maximum value at $\rho =\infty$ but the value is now
different:
\begin{equation}
\int\limits_{\textstyle 1}^{\textstyle 1/\sqrt{1-{\dot\rho}^2 (\infty)}}
\frac{dx}{\displaystyle 1-\frac{\kappa^2}{12\pi}\log x}=\frac{\E}{M}\quad .
\end{equation}
Like in (2), ${\dot\rho}(\infty)$ grows with $\E/M$ {\it but not
up to} {1}: 
\begin{equation}
{\dot\rho}(\infty)=1-\frac{1}{2}\exp\left(-\frac{24\pi}{\kappa^2}\right)
\; ,\qquad \frac{\E}{M}\to\infty
\end{equation}
{\it and this is the principal consequence of the vacuum backreaction.}

The insertion of (13) in (1) restores the conservation laws:
\begin{equation}
\frac{\Delta\E}{\E}=\frac{\Delta e}{e}=\frac{1}{2}+\kappa^2 O(1)\; ,
\qquad \frac{\E}{M}\to\infty\; .
\end{equation}
Up to 50\% of energy and charge can be extracted from a source by
raising its initial energy. But the main result is, of course, (11).
The vacuum doesn't change the electric potential as one could expect.
It changes {\it the kinematics of motion} like relativity theory does.
Furthermore, within a given type of coupling, this change is universal.
It does not depend on the parameters of the source, only on the
coupling constant $\kappa^2$. The vacuum appears as a medium in which
the velocity of light is less than $c$ but there is one special thing
about this medium: it cannot be escaped.

Like the vacuum radiation itself, its backreaction is a semiclassical 
effect, and the avoidance of the ultraviolet problem deserves a note.
This problem manifests itself in the fact that the distribution
$\e (t,r)$ that solves the expectation-value equations is singular
on the shell's surface:
\begin{equation}
\e (t,r)\biggl|_{r\to\rho (t)\pm 0}=\mp e\,\frac{\kappa^2}{24\pi}
\log |r-\rho (t)|
\end{equation}
(Fig. 2). Nevertheless, the force exerted on the shell is finite and
is unambigously obtained by making the sum (10) in the spectral integral.
This is equivalent to giving the shell a Compton width. The ultraviolet
problem concerns the spacelike vicinity of the shell but not its motion.
It is important that the infinite jump of $\e (t,r)$ across the shell
is constant. Owing to this fact, the fluxes across the shell are finite,
and so is the force of their backreaction.

Another object of worry is the vicinity of $r=0$ since
$\r=(e^2/2\E)$ while $\E$ increases. However, one can consider two
procedures of raising the energy: keeping $e$ fixed and keeping
$\r$ fixed. Only in the first case does the shell probe small scales
whereas the results above are the same in both cases. The details
of radiation at early time may be sensitive to the small scales but
the resultant distribution of charges at late time may not.

This work was done at the Princeton Institute for Advanced Study
which extended to the author its hospitality and support. The
appearance of Eq. (11) in Princeton is noteworthy. The work was 
supported in part also by the Russian Foundation for Fundamental
Research (Grant 99-02-18107) and INTAS (Grant 93-493-ext).
\vspace{15mm}
\hrule\smallskip
\begin{itemize}
\item[[1]] A.G. Mirzabekian and G.A. Vilkovisky, Phys. Lett.
B 414 (1997) 123; Ann. Phys. 270 (1998) 391 [gr-qc/9803006].
\item[[2]] G.A. Vilkovisky, hep-th/9812233 (to appear in Phys. Rev. D).
\end{itemize}

\newpage

\begin{center}
\section*{\bf Figure captions}
\end{center}
\begin{itemize}
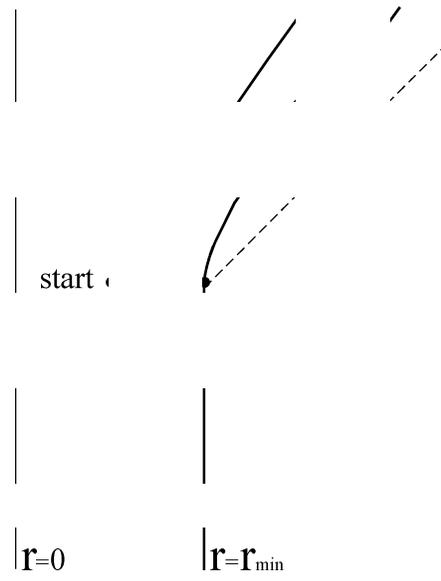
\item[Fig.1.] The world line of the shell on the $r,t$ plane. The
broken line is the outgoing light ray.
\item[Fig.2.] The distribution $\e (t,r)$ for a given $t$.
\end{itemize}
 
\newpage
\begin{figure}[p]
\begin{picture}(360,450)
\put(135,285){
\newbox\onebox
\newdimen\onew
\font\onea=onea at 72.27truept
\setbox\onebox=\vbox{\hbox{%
\onea\char0\char1\char2}}
\onew=\wd\onebox
\setbox\onebox=\hbox{\vbox{\hsize=\onew
\parskip=0pt\offinterlineskip\parindent0pt
\hbox{\onea\char0\char1\char2}
\hbox{\onea\char3\char4\char5}
\hbox{\onea\char6\char7\char8}
\hbox{\onea\char9\char10\char11}}}
\ifx\parbox\undefined
    \def\setone{\box\onebox}
\else
    \def\setone{\parbox{\wd\onebox}{\box\onebox}}
\fi
\setone}
\end{picture}
\begin{center}
{\LARGE Fig.1}
\end{center}
\end{figure}
\newpage
\begin{figure}[p]
\begin{picture}(360,450)
\put(90,270){
\newbox\twobox
\newdimen\twow
\font\twoa=twoa at 72.27truept
\setbox\twobox=\vbox{\hbox{%
\twoa\char0\char1\char2\char3}}
\twow=\wd\twobox
\setbox\twobox=\hbox{\vbox{\hsize=\twow
\parskip=0pt\offinterlineskip\parindent0pt
\hbox{\twoa\char0\char1\char2\char3}
\hbox{\twoa\char4\char5\char6\char7}
\hbox{\twoa\char8\char9\char10\char11}
\hbox{\twoa\char12\char13\char14\char15}}}
\ifx\parbox\undefined
    \def\settwo{\box\twobox}
\else
    \def\settwo{\parbox{\wd\twobox}{\box\twobox}}
\fi
\settwo}
\end{picture}
\begin{center}
{\LARGE Fig.2}
\end{center}
\end{figure}

\end{document}